\documentclass[twocolumn, 10pt, aps, amsmath, amssymb, floatfix, showpacs, prb, citeautoscript]{revtex4-2}

\usepackage{graphicx}
\usepackage[caption=false]{subfig}
\usepackage{amsmath}
\usepackage{amssymb}
\usepackage{bm}
\usepackage{xcolor}
\usepackage{upgreek}
\usepackage[colorlinks, citecolor={blue!80!black}, urlcolor={blue!80!black}, linkcolor={red!50!black}]{hyperref}
\usepackage{bookmark}
\usepackage[version=4]{mhchem}
\usepackage{microtype}
\usepackage{colordvi}

\newcommand{\comment}[1]{}

\renewcommand{\comment}{\paragraph}

\begin{document}
\title{Valley relaxation in a single-electron bilayer graphene
quantum dot}

\author{Lin Wang}
\affiliation{Department of Physics, University of Konstanz, D-78457 Konstanz, Germany}
\author{Guido Burkard}
\affiliation{Department of Physics, University of Konstanz, D-78457 Konstanz, Germany}

\begin{abstract}
We investigate the valley relaxation due to intervalley coupling in a single-electron 
bilayer graphene quantum dot. The valley relaxation is assisted by both the emission of acoustic phonons via the deformation potential and bond-length change mechanisms and $1/f$ charge noise. 
In the perpendicular magnetic-field dependence of the valley relaxation time $T_1$, we predict 
a monotonic decrease of $T_1$ at higher fields due to electron-phonon coupling, which is in good agreement with 
 recent experiments by Banszerus \emph{et al.}~\cite{Banszerus2024}. We find that the dominant 
valley relaxation channel in the high-field regime is the electron-phonon coupling via the deformation potential. At lower fields, we predict that a peak in $T_1$ can arise from the competition between $1/f$ charge noise and electron-phonon scattering due to bond-length change. We also find 
that the interlayer hopping $\gamma_3$ opens a valley relaxation channel for electric charge noise for rotationally symmetric quantum dots in bilayer graphene.
\end{abstract}

\maketitle

\section{Introduction}
Bernal-stacked bilayer graphene (BLG) has a tunable band gap controlled by an out-of-plane electric 
field~\cite{McCann2006,Konschuh85,McCann_2013, Min75, Castro99, Zhang2009, Nilsson78}. 
This semiconducting property enables the formation of quantum dots (QDs) in BLG via electrostatically-induced quantum confinement. 
The possibility of hosting electron spin qubits in graphene-based QDs \cite{Trauzette2007} has received some attention due to their expected long-lasting spin coherence 
 including low hyperfine interaction and weak spin-orbit coupling~\cite{Kane95, Huertas74, Min74, Yao75, Boettger75, Fischer80, Gmitra80, Huertas103, Abdelouahed82, Konschuh82}. Recently, long spin relaxation times of a 
single-electron state in BLG QDs exceeding $200\,\mu {\rm s}$ ~\cite{Banszerus13} and even up to $50 {\rm ms}$ ~\cite{Gachter3} were reported, suggesting that BLG is a promising material for spin qubits. 

In addition to spin, the valley pseudospin is another degree of freedom in graphene and other van der Waals materials, arising from their two-dimensional honeycomb 
lattice structure. While in the two-dimensional transition-metal dichalcogenides, the strong spin-orbit coupling locks the spin and valley degrees of freedom together, leading to interesting combinations of spin and valley qubits \cite{Brooks2020,Krishnan2023,Aliyar2024}, the weak spin-orbit coupling in graphene ensures that the spin and valley degrees of freedom are nearly independent.  Specifically, BLG has two independent energy valleys located at the ${\bm K}$ and ${\bm K^{\prime}}$ points 
of the hexagonal Brillouin zone. With broken spatial inversion symmetry, the two inequivalent valleys in gapped BLG experience 
opposite Berry curvatures and associated orbital magnetic moments \cite{Knothe98,Eich2018,Banszerus2020}. This leads to a valley splitting which grows linearly in the applied  
out-of-plane magnetic field, and which can be viewed as a valley Zeeman splitting similar to the spin Zeeman splitting. The valley Zeeman effect has already been 
demonstrated recently by single-carrier measurements in BLG QDs \cite{Banszerus2021}. This effect provides a promising path towards 
controlling the valley degree of freedom and to further establish valley-based electronics (valleytronics) \cite{Rycerz2007,Schaibley2016} as well as valley-based qubits in graphene QDs \cite{Rohling2012,Rohling2014}.
To assess the potential of valley bits and qubits, the valley relaxation time of single-electron states in a BLG 
QD is a crucial parameter since it limits the lifetime of both the encoded classical and quantum information. Recently, Banszerus~\emph{et al.} 
measured valley relaxation times as large as several microseconds using pulsed-gate spectroscopy~\cite{Banszerus2024}. In the out-of-plane 
magnetic filed dependence of valley relaxation time, a monotonic decay at higher fields and  a peak at smaller fields was observed. To explain the experiment, theoretical work on valley relaxation in BLG QDs is required.

\begin{figure}
\includegraphics[width=1.0\columnwidth]{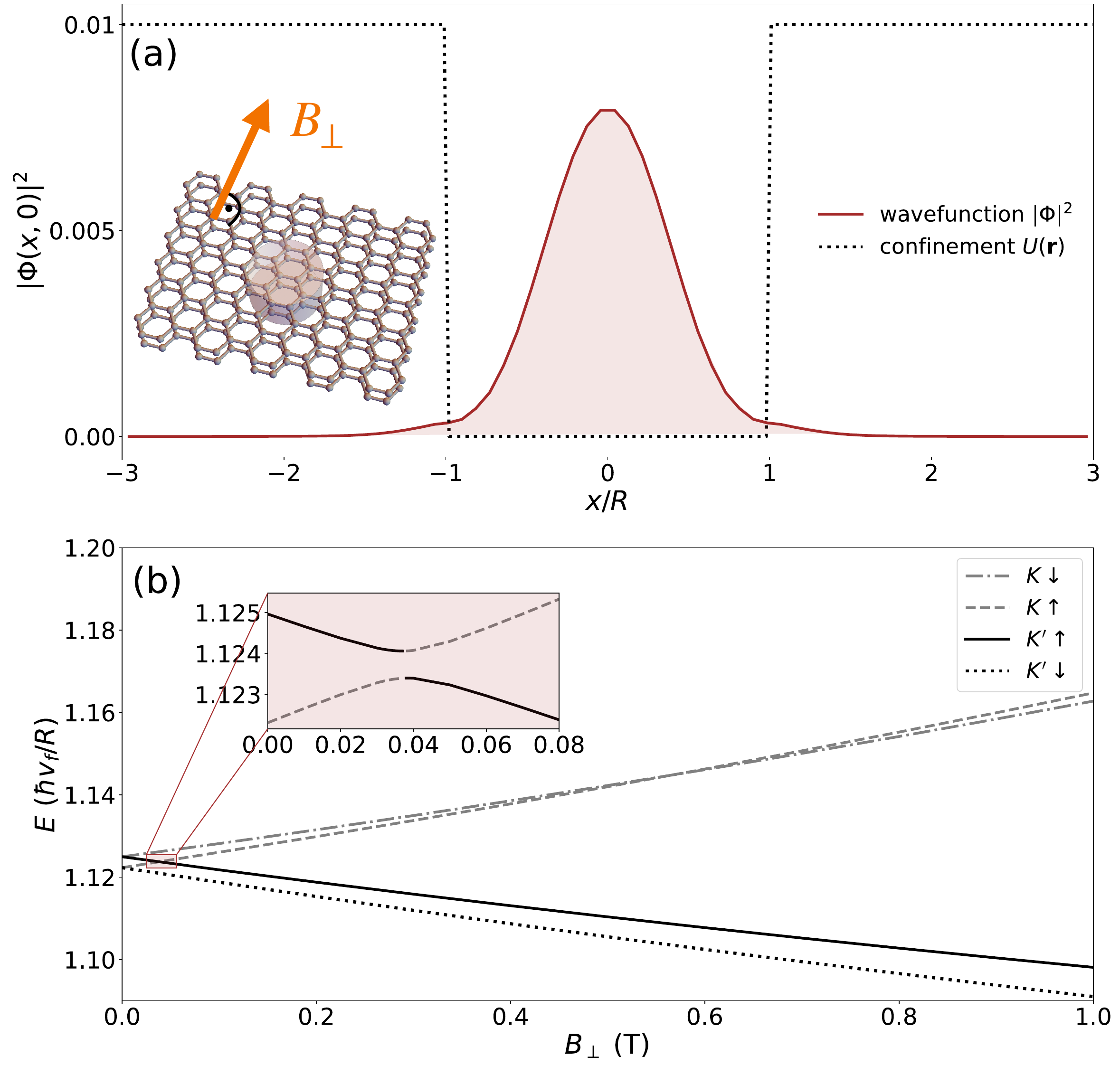}
\caption{(a) Calculated electronic probability density $|\phi(x)|^2$ (brown solid curve) of the lowest energy level at $B_{\perp}=0.2\ $T as a function of position  $x$ for fixed $y=0$. An electrostatic step potential (black dotted curve) defines a QD of radius $R$. Inset: schematic picture of a BLG QD. (b) Bound-state energy levels as a function of perpendicular magnetic field $B_{\perp}$ in a BLG QD. The energy splitting between different valleys is dominated by the valley Zeeman effect with valley $g$-factor $g_v=31$. The anticrossing between the bound states ${\bm K}\uparrow$ and ${\bm K^{\prime}}\uparrow$ is enlarged in the inset. For this calculation, we have set
$U_0=42.5\ $meV, $V=25\ $meV, $R=25\ $nm, $\Delta_{\rm i}=43\ \mu$eV, and $\Delta_{\rm KK^{\prime}}=10\ \mu$eV.
Energies are plotted in units of $\hbar v_f/R$ where $v_f$ denotes the Fermi velocity, see Table~\ref{tab:kp_mu}.}
\label{fig:energy_levels}
\end{figure}
In this paper, we investigate both the bound-state energy levels and 
the valley relaxation time of a single-electron BLG QD 
as a function of the out-of-plane magnetic fields $B_\perp$ (see Fig.~\ref{fig:energy_levels}).
We employ an exact diagonalization method to obtain the energy levels and the eigenstates. 
At zero magnetic field, the lowest four energy levels form 
two Kramers pairs separated by an intrinsic spin-orbit gap. At finite magnetic fields, the energy levels become linearly 
dependent on $B_\perp$ due to both spin and valley Zeeman effects. At the value of $B_\perp$ where  the intrinsic spin-orbit gap and valley Zeeman energy splitting coincide,  intervalley scattering causes an anticrossing between the two different valley states. 
Based on the single-electron spectrum of the BLG QD, we then calculate the valley relaxation time using Fermi's golden rule. The valley relaxation channels are enabled by the intervalley coupling together with (i) $1/f$ charge noise and (ii) electron-phonon coupling via the deformation potential and the bond-length change. We find a peak at lower fields and a monotonic decay at higher fields. The low-field peak arises from the competition between the contributions of $1/f$ 
charge noise and bond-length change electron-phonon coupling. The monotonic decay at higher fields is due to the dominant contribution of the deformation potential. Moreover, we find that the interlayer hopping $\gamma_3$ opens a valley relaxation channel for electric charge noise when the quantum dot has rotational symmetry. Upon detailed comparison with the experiment \cite{Banszerus2024}, we find a good agreement between experiment and theory at higher magnetic fields.

The remainder of this paper is organized as follows. In Section~\ref{sec:model} we describe our model and the methods used for the calculation of the valley relaxation time. In Sec.~\ref{sec:mechanisms} we describe the relevant mechanisms that contribute to valley relaxation, and in Sec.~\ref{sec:results} we show the resulting valley  relaxation times obtained from the numerical evaluation of our model. Finally, we conclude our work in Sec.~\ref{sec:conclusions}.

\section{Model and Method}
\label{sec:model}
We consider Bernal (AB) stacked BLG in the presence of a homogeneous out-of-plane magnetic field $B_{\perp}$ and an electrostatic confinement potential 
$U({\bf r})$ with ${\bf r}=(x,y)$. The total single-particle Hamiltonian can be written as 
\begin{eqnarray}
H_{\rm QD}=H^{\tau}({\bf k})+U({\bf r})+H_{\rm SO}+H_{\rm Z}+H_{KK'},\label{total_hamil}  
\end{eqnarray}
where $H^{\tau}({\bf k})$ is an effective $4\times 4$ Hamiltonian describing the spinless $\pi$-band structure of BLG near the ${\bm K}$ 
and ${\bm K^{\prime}}$ valleys \cite{Konschuh85,Wang87},
\begin{eqnarray}
H^{\tau}({\bf k})=\left(\begin{array}{cccc}
V & \gamma_0 p & \gamma_4 p^* & \gamma_1 \\
\gamma_0 p^* & V & \gamma_3 p & \gamma_4 p^* \\
\gamma_4 p & \gamma_3 p^* & -V & \gamma_0 p \\
\gamma_1 & \gamma_4 p & \gamma_0 p^* & -V
\end{array}\right),\label{eq5}
\end{eqnarray}
in the on-site orbital Bloch basis $\Psi_{\rm A_1}({\bf k})$, $\Psi_{\rm B_1}({\bf k})$,
$\Psi_{\rm A_2}({\bf k})$ and
$\Psi_{\rm B_2}({\bf k})$, where A$_1$ and B$_1$ refer to the A and B sublattices in the lower layer, A$_2$ and B$_2$ denote the A and B sublattices in the upper layer, and ${{\bf k}}=(k_x,k_y)$ represents the two-dimensional
wave vector measured from the ${\bm K}$ ($\tau=1$) or ${\bm K^{\prime}}$ ($\tau=-1$) point.
Here, $2V$ describes the potential difference between the two graphene layers, which is controlled by an out-of-plane electric field, and
$\gamma_0$ and $\gamma_1$ represent the nearest-neighbor intralayer and
interlayer hoppings whereas $\gamma_3$ and $\gamma_4$ are the indirect hopping parameters between the two layers.
The momentum dependence is given by $p({\bf k})=-\sqrt{3}a(\tau {k}_x-i{k}_y-ixB_{\perp}e/2-\tau yB_{\perp}e/2)/2$ which includes the orbital effect due to the out-of-plane magnetic 
field, with $a$ being the lattice constant. The second term in Eq.~\eqref{total_hamil} is the confinement potential, which for simplicity is chosen to be a finite circularly symmetric step potential,
\begin{eqnarray}
U({\bf r})=\left\{\begin{array}{cc}
U_0,     &r\ge R,  \\
0,     &r<R, 
\end{array}\right.
\end{eqnarray} 
where $U_0$ and $R$ denote the potential depth and QD radius, respectively. The third term 
$H_{\rm SO}=\Delta_{\rm i}\tau\sigma_z s_z+\Delta_{\rm R}(\tau\sigma_x s_y-\sigma_y s_x)$ describes 
the spin-orbit coupling with strength $\Delta_{\rm i}$ ($\Delta_{\rm R}$) for the intrinsic (Rashba) spin-orbit effects, and $\sigma_{x,y,z}$ ($s_{x,y,z}$) are the Pauli matrices for the sublattice (spin) degree of freedom. 
While the intrinsic spin-orbit coupling does not contribute to the inter-valley mixing, it does have an effect on the spin and valley resolved QD energy levels. Note however  that the contribution of the Rashba term is negligible and therefore we take $\Delta_{\rm R}$ to be zero 
throughout this paper. The fourth term describes the spin Zeeman coupling as $H_{\rm Z}=g_s\mu_B s_zB_{\perp}/2$ with $g_s$ the 
spin $g$-factor and $\mu_{B}$ the Bohr magneton. The last term $H_{KK'}=\Delta_{KK^{\prime}}\tau_x/2$ represents the 
intervalley coupling possibly induced by disorder where $\Delta_{KK^{\prime}}$ quantifies the intervalley coupling strength and $\tau_x$ denotes the Pauli $x$ matrix for valley~\cite{Banszerus2021,Klinovaja86}. All the parameters are listed in Table~\ref{tab:kp_mu}.
\begin{table}[t]
\caption{Parameters used in the calculation: $\gamma_{0,1,3,4}$, $g_s$, and $a$ are introduced in Eq.~(\ref{total_hamil}), $v_{\rm TA,LA}$ and $\rho$ are used in Eq.~(\ref{ele_phonon}), and $v_f$ appears in Fig.~\ref{fig:energy_levels}.}
\begin{tabular}{c c  c | c c}
\hline\hline
$\gamma_0$ & $2.6\ $eV &\quad\quad & $\gamma_1$ & $0.339\ $eV \\
$\gamma_3$ & $0.28\ $eV & & $\gamma_4$ & $-0.14\ $eV \\
$g_s$ & $2$ & & $a$ & $2.46\ \textup{\AA}$\\
$v_{\rm TA}$ & $1.22\times 10^4\ $m/s & & $v_{\rm LA}$ & $1.95\times 10^4\ $m/s\\
$v_f$ & $8\times 10^5\ $m/s & & $\rho$ & $1.52\times 10^{-7}\ $g/cm$^{-2}$\\
\hline\hline
\end{tabular}
\label{tab:kp_mu}
\end{table}

Note that in the absence of the Rashba term, the total Hamiltonian (\ref{total_hamil}) can be divided into two 
independent spin blocks, i.e., spin up and spin down. Within each spin block, we first solve the Schr\"{o}dinger equation 
of the Hamiltonian $H_0=H^{\tau}({\bf k})+U({\bf r})$ numerically by discretizing two-dimensional real space using a square 
lattice grid. To rid ourselves of the Fermi doubling problem arising from the lattice discretization, we include a Wilson mass 
term $wk^2$ in $H_0$ with $w$ denoting the Wilson mass~\cite{Messias96}. Then, we take into account the remaining terms in Eq.~(\ref{total_hamil}) in the eigenbasis of $H_0$ in 
both valleys. Finally, we can exactly diagonalize the total Hamiltonian and obtain single-particle eigenvalues and eigenfunctions  which we plot in Fig.~\ref{fig:energy_levels}(b).

\section{Valley relaxation mechanisms}
\label{sec:mechanisms}
With the eigenvalues and eigenstates at  hand, we can now calculate the valley relaxation rate from the initial state $|i\rangle$ 
to the final state $|f\rangle$ using Fermi's golden rule. Here, we consider valley relaxation due to (i) the electron-phonon 
coupling via the deformation potential and the bond-length change mechanisms \cite{Struck2010,Droth2011,Droth2013}, and (ii) $1/f$ charge noise \cite{Hosseinkhani2021,Hosseinkhani2022}.

\subsection{\footnotesize Relaxation induced by electron-phonon coupling}
We consider two different electron-phonon interaction mechanisms, the deformation potential on the one hand and the
bond-length change on the other. The deformation potential is induced by a phonon-induced area change in the unit cell, whereas the bond-length change is caused by a modified hopping matrix element. Since we are interested in the low-energy regime, only acoustic 
phonons are considered. Furthermore, out-of-plane phonon modes are irrelevant since we assume that the host BLG sheet is placed on a substrate. Therefore, we  take into account in-plane longitudinal-acoustic (LA) and transversal-acoustic (TA) 
modes only. The electron-phonon coupling is described in the sublattice basis as \cite{Ando2005},
\begin{align}
H_{\rm EPC}^{\lambda q}=\frac{q}{\sqrt{A\rho\Omega_{{\bf q},\lambda}}}\left(\begin{array}{cc}
g_1a_1     &g_2a^*_2  \\
g_2a_2     &g_1a_1 
\end{array}\right)(e^{i{\bf q}\cdot{\bf r}}b_{\lambda q}^{\dagger}-e^{-i{\bf q}\cdot{\bf r}}b_{\lambda q}),
\label{ele_phonon}
\end{align}
with $A$ the area of the graphene sheet, $\rho$ the mass density of BLG, $g_1$ ($g_2$) the coupling strength of the deformation 
potential (bond-length change), $a_1=i$ and $a_2=ie^{2i\phi_{\bf q}}$ for LA phonons, and $a_2=e^{2i\phi_{\bf q}}$ and 
$a_1=0$ for TA phonons. The phonon energy is given by $\Omega_{{\bf q}, \lambda}=v_{\lambda}q$ with $v_{\lambda}$ the sound velocity for the phonon branch $\lambda={\rm TA},{\rm LA}$, while $b_{\lambda q}^\dagger$ and $b_{\lambda q}$ denote the creation and annihilation operators for branch $\lambda$ phonons with wavevector $q$. 

Using Fermi's golden rule, we can calculate the valley relaxation rate due to electron-phonon coupling from $|i\rangle$ with energy $\epsilon_i$ to $|f\rangle$ with energy $\epsilon_f$ as 
\begin{equation}
 \frac{1}{T_1}=2\pi A\sum_{\lambda}\int\frac{d^2q}{(2\pi)^2}|\langle i|H_{\rm EPC}^{\lambda q}|f\rangle|^2\delta(\epsilon_f-\epsilon_i+\Omega_{{\bf q},\lambda}).
\end{equation}
Note that only the phonon emission process is taken into account here by assuming that the temperature is much lower than the valley splitting.

\subsection{\footnotesize Relaxation induced by \texorpdfstring{$1/f$}{} charge noise}
Electric (charge) noise with its typical $1/f$ power spectral density often arises as a consequence of fluctuating two-level systems in the environment of the localized QD electron. More precisely, the electric charge noise 
spectra can be given by $S_{\rm E}(\omega)=S_0/\omega^{\alpha}$ where $S_0$ stands for the 
power spectral density at $1\ $Hz and the exponent $\alpha$ is device dependent and typically 
reported to be between $0.5$ and $2$~\cite{Kranz32}. We can calculate the valley relaxation rate from $|i\rangle$ to $|f\rangle$ using 
\begin{eqnarray}
\frac{1}{T_1}&=&\frac{4\pi e^2}{{\hbar}^2}S_{\rm E}(\epsilon_i-\epsilon_f)\sum_j|\langle i|r_j|f\rangle|^2\\
             &=&\frac{4\pi e^2}{{\hbar}^2}S_{\rm E}(\epsilon_i-\epsilon_f)\frac{|\langle i|r_+|f\rangle|^2+|\langle i|r_-|f\rangle|^2}{2},\label{noise} 
\end{eqnarray}
with ${\bf r}=(x, y)$ and $r_{\pm}=x\pm iy$.

In the absence of $\gamma_3$ in \eqref{eq5}, the Hamiltonian of each valley has rotational symmetry and hence
commutes with the total angular-momentum operator $J_z=L_z+\tau\hbar(\sigma_z/2-\eta_z/2)$. Here, $L_z$ and $\eta_z$ represent the
orbital angular momentum operator and the Pauli $z$ matrix for the layer degree of freedom, respectively. Thus, we have $J_z\Phi(\tau, m)=m\hbar\Phi(\tau,m)$ where 
$\Phi(\tau, m)$ is an eigenstate within each valley $\tau=\pm 1$. Due to the existence of the intervalley coupling $H_{KK'}$, the states 
$\Phi(\tau=1, m_1)$ in the $K$ valley can be coupled to the states $\Phi(\tau=-1, m_2)$ in the $K^{\prime}$ valley with coupling matrix element
$M^{m_1m_2}_{KK^{\prime}}=\Delta_{KK^{\prime}}\Phi^{\dagger}(\tau=1,m_1)\Phi(\tau=-1,m_2)/2$. It is easy to demonstrate that 
$M^{m_1m_2}_{KK^{\prime}}$ is nonzero only when $m_1=m_2$. This means that only the states with the same angular momentum 
in two valleys are coupled. Further, by referring to Eq.~(\ref{noise}), the valley relaxation 
due to $1/f$ noise is absent unless the angular momentum between the initial and final states differs by $\pm 1$. This 
can be used as a selection rule for valley relaxation due to $1/f$ noise in the presence of rotational symmetry. 
This selection rule can be generalized to other types of electric charge noise.

\section{Numerical results}
\label{sec:results}
In Fig.~\ref{fig:energy_levels}(a), we show a schematic picture of a BLG QD defined by an electrostatic step potential. The probability density associated with the lowest energy level at $B_{\perp}=0.2\ $T is plotted as a function of the coordinate $x$ for fixed $y=0$. This indicates that the wavefunction is localized within the QD. We then calculate the lowest four bound-state energy levels as a function of out-of-plane 
magnetic field, shown in Fig.~\ref{fig:energy_levels}(b) with labels $|K\uparrow\rangle$, $|K\downarrow\rangle$, 
$|K^{\prime}\uparrow\rangle$ and $|K^{\prime}\downarrow\rangle$. At zero magnetic field, we find 
two Kramers pairs ($|K\uparrow\rangle$, $|K^{\prime}\downarrow\rangle$) and ($|K\downarrow\rangle$, $|K^{\prime}\uparrow\rangle$), 
separated by the intrinsic spin-orbit gap of around $70\ \mu$eV. At finite magnetic fields, all of these energy levels show a linear 
dependence with slopes  $\frac{1}{2}(\pm g_s\pm g_v)\mu_B$ according to the spin and valley Zeeman 
effects. Here, we find $g_v=31.0$ as the valley $g$-factor. In addition, we find an anticrossing between the states $|K\uparrow\rangle$ 
and $|K^{\prime}\uparrow\rangle$, which is enlarged in the inset of Fig.~\ref{fig:energy_levels}(b). This anticrossing results from the intervalley coupling $H_{KK'}$.

\begin{figure}[t]
\includegraphics[width=1.0\columnwidth]{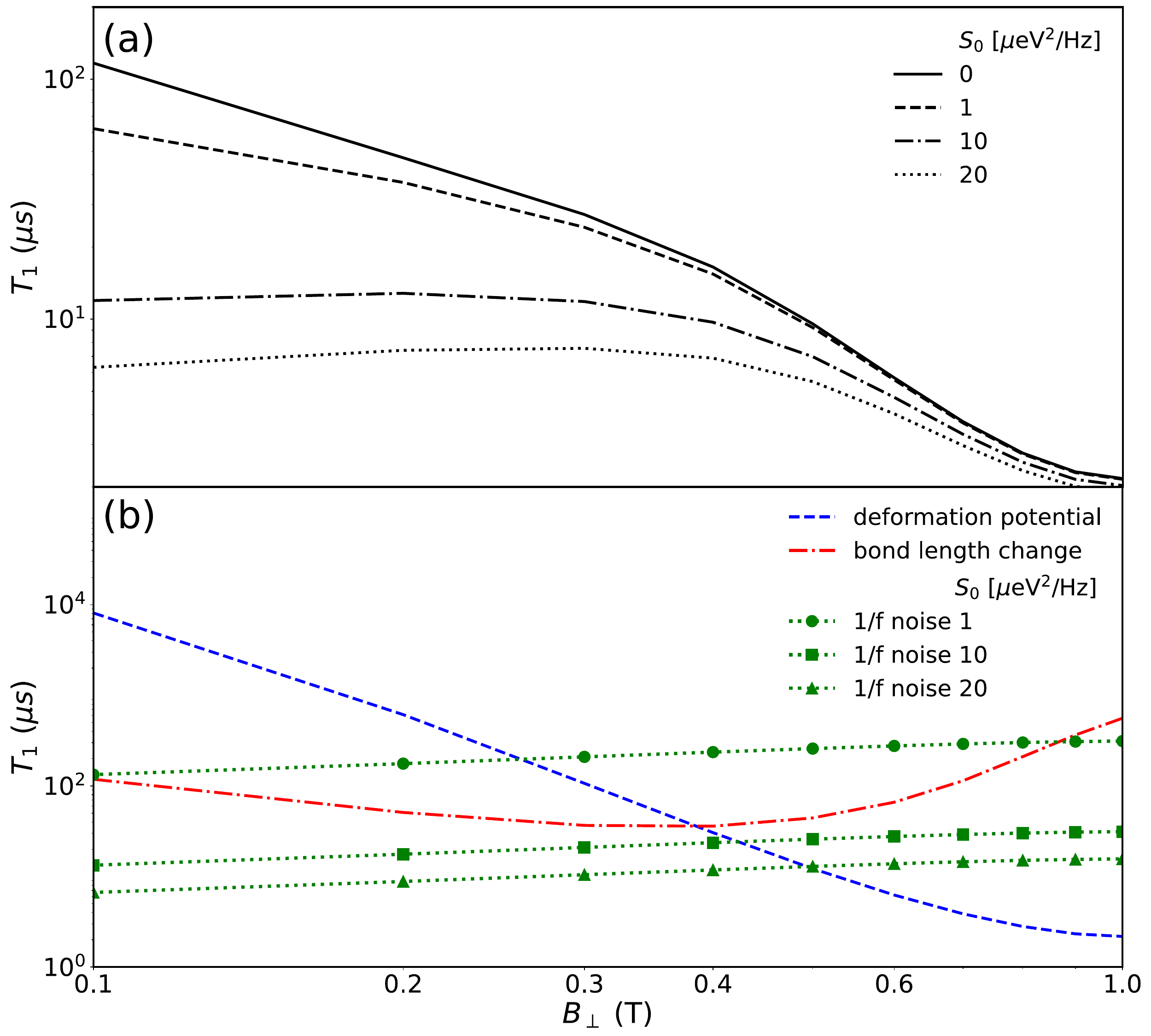}
\caption{(a) Total valley relaxation time $T_1$ as a function of perpendicular field $B_{\perp}$ on a double-logarithmic scale, for different noise strengths $S_0=0, 1, 10, 20\ \mu$eV$^2/$Hz ($1/f$ noise). (b) Valley relaxation time due to deformation potential (blue dashed curve), bond-length change (red dotted-dashed curve) and $1/f$ 
charge noise (green dotted curves). For the $1/f$ charge noise, $S_0=1, 10, 20\ \mu$eV$^2/$Hz are represented by dots, squares, and triangles, respectively. In the calculation, we have used $U_0=42.5\ $meV, $V=25\ $meV,  $R=25\ $nm, $\Delta_{\rm i}=43\ \mu$eV, $\Delta_{KK^{\prime}}=10\ \mu$eV, $g_1=50\ $eV, $g_2=2.8\ $eV, and $\alpha=0.5$.} 
\label{fig:different_S0}
\end{figure}

\subsection{Valley relaxation: theoretical calculation}
As mentioned previously, in the absence of the Rashba spin-orbit term, our system is divided into two independent sectors, one with spin up and another with spin down. Then, the valley relaxation time 
can be calculated within each spin sector. Since the valley relaxation times of spin up and spin down are
very close to each other, we focus on the spin down sector in the present work.  

In Fig.~\ref{fig:different_S0}(a), the total resulting valley relaxation time is plotted as a 
function of the perpendicular magnetic field $B_\perp$ for different values of $S_0$ characterizing the strength of  the $1/f$ noise. In the absence of $1/f$ charge noise, 
the valley relaxation time shows a monotonic decrease with increasing magnetic field. With finite $1/f$ charge noise, 
the valley relaxation time becomes much shorter at lower fields, indicating that $1/f$ charge noise plays a 
more important role at lower fields. In addition, a peak is predicted at lower fields for sufficiently strong $1/f$ noise (large $S_0$). To understand these behaviors, we show the contributions of deformation potential, 
bond length change, and $1/f$ charge noise separately in Fig.~\ref{fig:different_S0}(b). In the absence of 
$1/f$ charge noise, the valley relaxation is dominated by the bond length change (deformation potential) at lower (higher) fields. 
Both the valley relaxation time due to the bond length change at lower fields and the deformation potential at 
higher fields decrease with  increasing magnetic field, arising from the increase of the valley splitting, i.e., the energy splitting between initial and final states. This gives rise to a monotonic decrease of valley relaxation time. When $1/f$ charge noise comes into 
play, the valley relaxation at higher fields is still dominated by the deformation potential. However, at lower fields, 
there exists a competition between the $1/f$ noise and bond-length change when the noise spectral density is strong. This competition leads to a peak in the magnetic field dependence.


In addition, we also investigate the valley relaxation with different exponents $\alpha$ in the $1/f$ charge noise power spectral density. We find that with increasing $\alpha$, the 
valley relaxation time exhibits an overall increase. In particular, the increase at lower fields is dramatic, suggesting 
the importance of $1/f$ charge noise at lower fields. Similar to the situation of different noise spectral densities mentioned above, a peak is also observed in the magnetic field dependence for 
small $\alpha$, attributed to the competition between $1/f$ charge noise and the bond length change mechanism.

\begin{figure}[t]
\includegraphics[width=1.0\columnwidth]{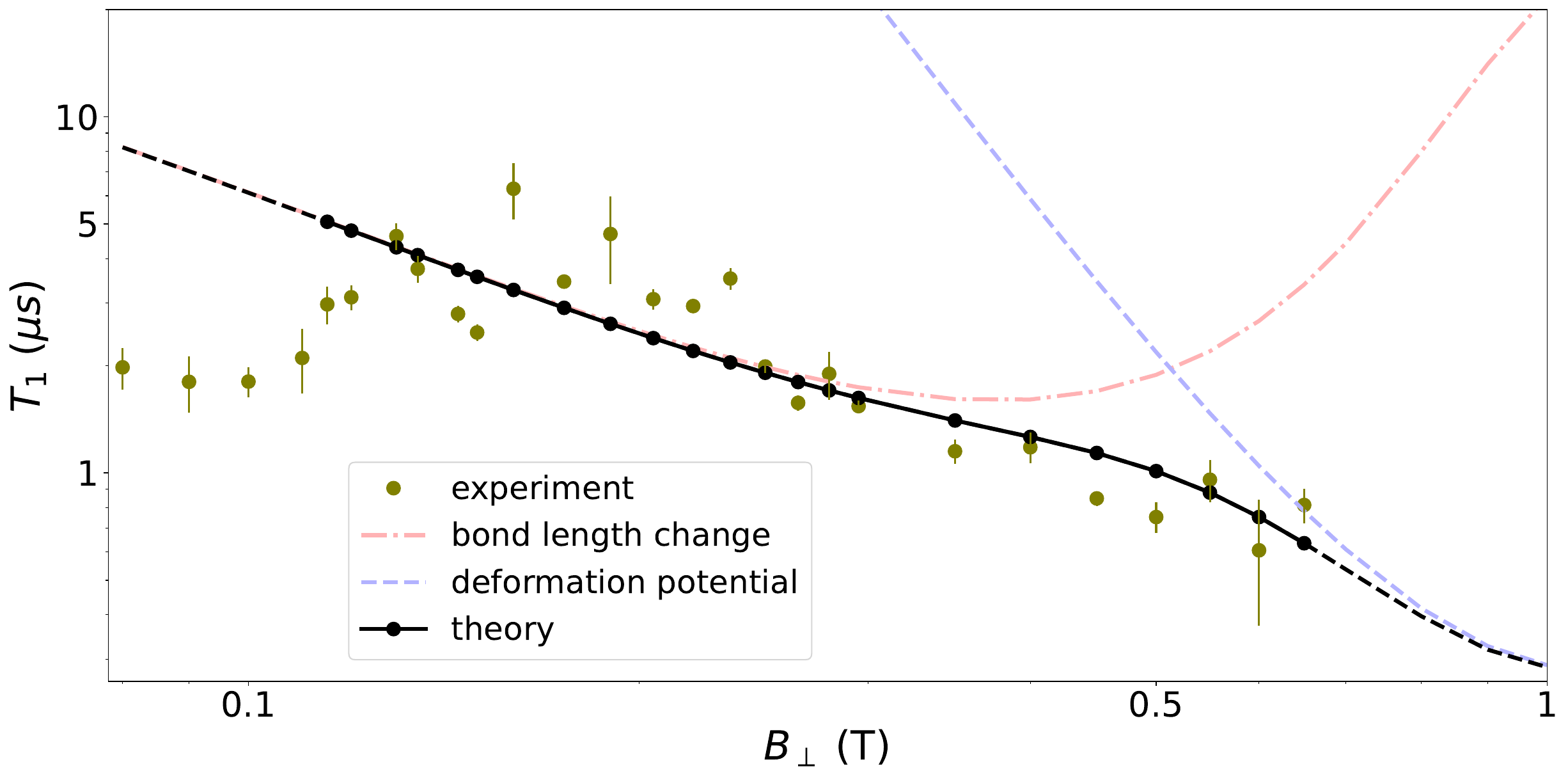}
\caption{Log-log plot of the valley relaxation time $T_1$ as a function of perpendicular magnetic field 
$B_{\perp}$. The olive dots stand for the experimental data from~\cite{Banszerus2024}, while the black curve with dots represents
the theoretical result obtained by fitting to the experimental data using least square method with fitting parameters $g_1=50\ $eV, $g_2=5.4\ $eV, $\Delta_{KK^{\prime}}=50\ \mu$eV, $R=25\ $nm, $U_0=39.6\ $meV, and $\Delta_{\rm i}=35\ \mu$eV. The blue dashed curve (red dotted-dashed curve) denotes the valley relaxation time due to deformation potential (bond-length change).}
\label{fig:T1_exp_total}
\end{figure}

\subsection{Comparison with  experiment}
Very recently, Banszerus~\emph{et al.}~\cite{Banszerus2024} reported experimental single-particle valley relaxation times in a BLG QD as a function of perpendicular magnetic field, 
shown in Fig.~\ref{fig:T1_exp_total} as olive dots. At higher fields, a monotonic decay is observed. 
To explain this behavior, we take into account the contribution of deformation potential and bond length change, while $1/f$ charge noise is not included due to its negligible contribution at higher fields. We perform a least square fit of the electron-phonon coupling strengths $g_1$ and $g_2$ to the measurement data, taking into account the experimental error bars . 
The result of our numerical fit is shown as the black curve with dots, which agrees with the experimental data both qualitatively and quantitatively. At lower fields, a peak is 
observed in the experimental data. Our theory suggests that the origin of this peak lies in a competition between electric noise and phonons.  However, similar peaks may also arise from a competition between different electron-phonon coupling mechanisms or from a valley crossing.  Further research can shed more light on the origin of the non-monotonic behavior of $T_1(B_\perp)$.


\section{Conclusions and Discussion}
\label{sec:conclusions}
We have studied the electronic valley relaxation time $T_1$ in a BLG QD with magnetic field $B_\perp$
perpendicular to the graphene plane. The valley relaxation is induced by the intervalley coupling together 
with $1/f$ charge noise and electron-phonon scattering via deformation potential and bond-length 
change. In the magnetic-field dependence of valley relaxation time, a peak at lower fields and a monotonic decay at higher fields are predicted, which agree with a recent experiment \cite{Banszerus2024}. The 
origin of the peak at lower fields is explained as arising from the competition between the contributions of $1/f$ noise and phonon emission via
bond-length change. The monotonic decay for large $B_\perp$ is due to the contribution of both the
deformation potential and bond length change at higher fields. In addition, we show that while the contribution of $1/f$ charge noise to valley 
relaxation is more important at lower fields due to smaller noise frequency, at higher fields the deformation potential dominates the valley relaxation process. 

The valley relaxation has also been studied in a BLG device containing gate-defined double QDs where the valley relaxation time between the valley triplet and singlet states was reported to be remarkably long~\cite{Garreis2024}. This experiment requires future research to understand the singlet-triplet valley relaxation mechanisms.

\acknowledgements
We would like to acknowledge P. M. Mutter for fruitful discussions.

\bibliographystyle{apsrev4-2}
\bibliography{valley_T1}
\end{document}